\documentclass[9pt,twocolumn,twoside]{revtex4-1}

\usepackage{graphicx}
\usepackage{dcolumn}
\usepackage{bm}
\usepackage{natbib}
\usepackage{physics}

\newcommand{\bea}{\begin{eqnarray}}
\newcommand{\eea}{\end{eqnarray}}

\begin{document}
\title{Optical Detection of Paramagnetic Defects in a CVD-grown Diamond}

\author{C. Pellet-Mary$^{1}$}
\author{P. Huillery $^{1}$}
\author{M. Perdriat $^{1}$}
\author{A. Tallaire $^{2}$}
\author{G. H\'etet$^{1}$} 

\affiliation{$^1$Laboratoire de Physique de l'Ecole normale sup\'erieure, ENS, Universit\'e PSL, CNRS, Sorbonne Universit\'e, Universit\'e Paris-Diderot, Sorbonne Paris Cit\'e, Paris, France.}
\affiliation{$^2$IRCP, Ecole Nationale Supérieure de Chimie de Paris, 11, rue Pierre et Marie Curie, 75005 Paris, France}

\begin{abstract}
The electronic spins of the nitrogen-vacancy centers (NV centers) in Chemical-Vapor-Deposition (CVD) grown diamonds form ideal probes of magnetic fields and temperature, as well as promising qu-bits for quantum information processing. 
Studying and controlling the magnetic environment of NV centers in such high purity crystals is thus essential for these applications.
We demonstrate optical detection of paramagnetic species, such as hydrogen-related complexes, in a CVD-grown diamond.
The resonant transfer of the NV centers' polarized electronic spins to the electronic spins of these species generates conspicuous features in the NV photoluminescence by employing magnetic field scans along the [100] crystal direction. 
Our results offer prospects for more detailed studies of CVD-grown processes as well as for coherent control of the spin of novel classes of hyper-polarized paramagnetic species.
\end{abstract}

\maketitle

The electronic spin properties of the negatively charged nitrogen-vacancy (NV$^-$) center in diamond has given rise to a wealth of applications in nanoscale sensing \cite{Rondin_2014} and quantum information science  \cite{DOHERTY20131}.
One major reason is that it can be optically polarized and read-out and features long coherence time even at ambient conditions. 
In order to optimize the capabilities of the NV center, its magnetic environment must be very well controlled. 
Synthetic growth of diamond crystals is now reaching a level of maturity that makes the most pristine diamond crystals almost flawless, thus eliminating the source of magnetic noise from other nearby impurities than the  NV sensor. The standard growth technique works through Chemical Vapor Deposition (CVD) of carbon atoms from a methane gas in an ultra-high vacuum environment. It offers the possibility to use crystals with 99.9\% purity, as well as using $^{12}$C enriched methane, hence removing spin fluctuations from $^{13}$C atoms and enhancing further the sensing capabilities of NV centers \cite{Achard}.

Electron-Paramagnetic-Resonance (EPR) spectroscopy is the method of choice for controlling the concentration of paramagnetic defects in materials.
Thanks to their high resolution, EPR spectrometers are also essential tools for understanding the conformation of many defects that remain in
CVD-grown diamond materials \cite{newton_epr_2007, glover_hydrogen_2004, glover_hydrogen_2003}. However, compared to confocal microscopy, this equipment is rather bulky and less cost-effective. 
It would also be ideal if these defects were detected with the very same technology that is employed for nanoscale sensing \cite{Rondin_2014} and quantum information processing \cite{Yao}. 
It would indeed offer the opportunity to hyper-polarize these defects using the optically polarized NV center and to employ them as quantum bits at ambient conditions.

 \begin{figure}[htbp]
\centering
{\includegraphics[width=\linewidth]{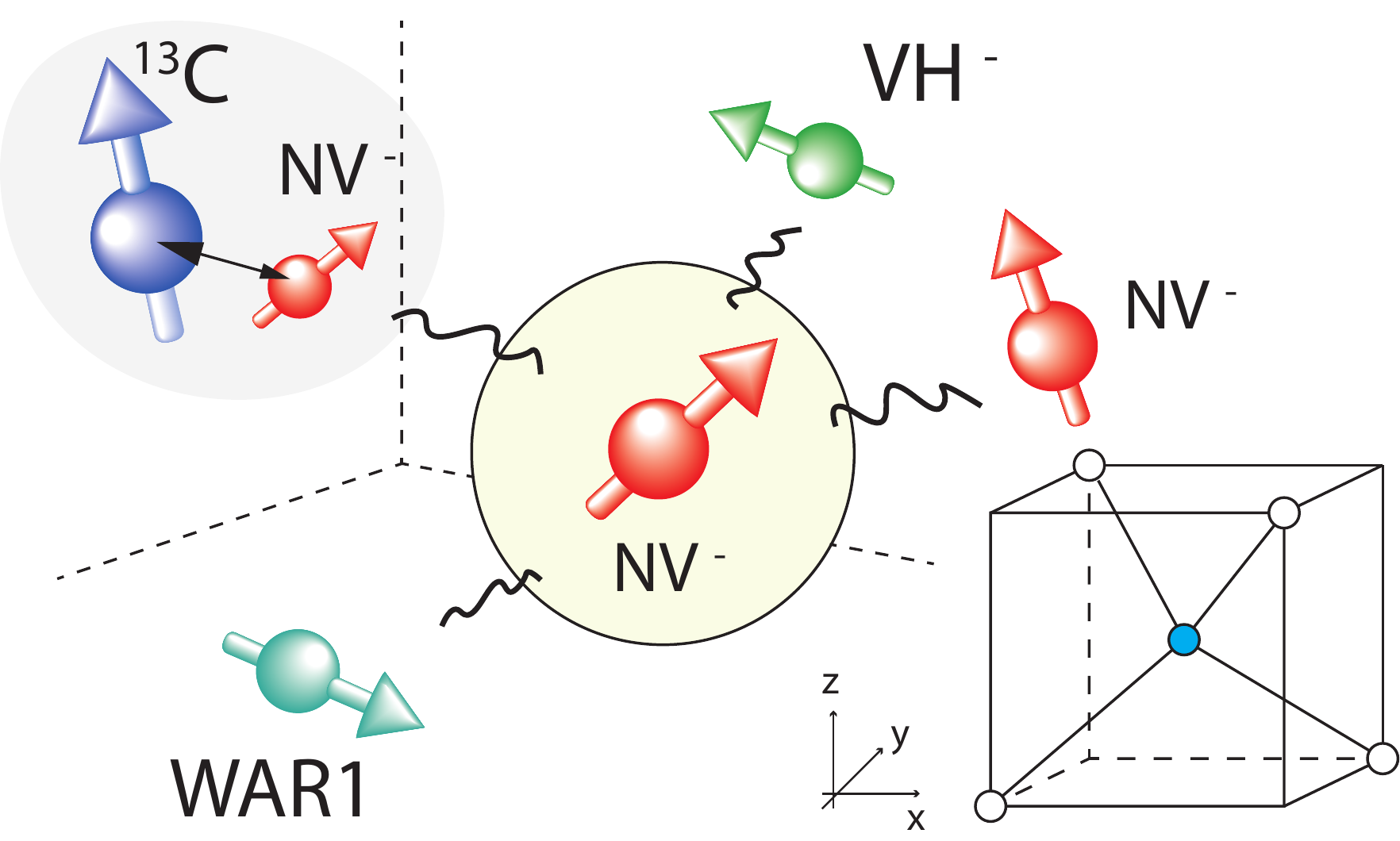}}
\caption{Schematics showing various paramagnetic defects in CVD-grown diamonds interacting with the electronic spin of a negatively charged nitrogen-vacancy center. {Bottom right}: Atomic positions in the diamond unit cell.}
\label{fig1}
\end{figure}



Here, we use confocal laser microscopy to detect paramagnetic defects {\it via} the coupling to a high-density NV spin ensemble in a CVD-grown diamond.
Fig. \ref{fig1} shows a schematics of the various paramagnetic defects that have been coupled to the nitrogen-vacancy center and detected optically in our study.
The detection is realized by measuring the NV photoluminescence while performing magnetic field scans to resonantly enhance dipole-dipole interactions and observe cross-relaxations (CR).
Cross-relaxation typically takes place when the electronic or nuclear spins of two atomic species exchange their polarizations via resonant magnetic dipole-dipole interactions. 
If the spin of an NV center A is polarized and coupled to an unpolarized spin B with a much larger relaxation rate, it will lose part of its polarization at the expense of B and thus see a drop of its photoluminescence (PL) rate. 
Tuning the frequency of both spin transitions so that are co-resonant will result in a reduction of the PL of NV A, thus enabling detection of the spin energy of B.

\begin{figure}[ht]
\centering
{\includegraphics[width=\linewidth]{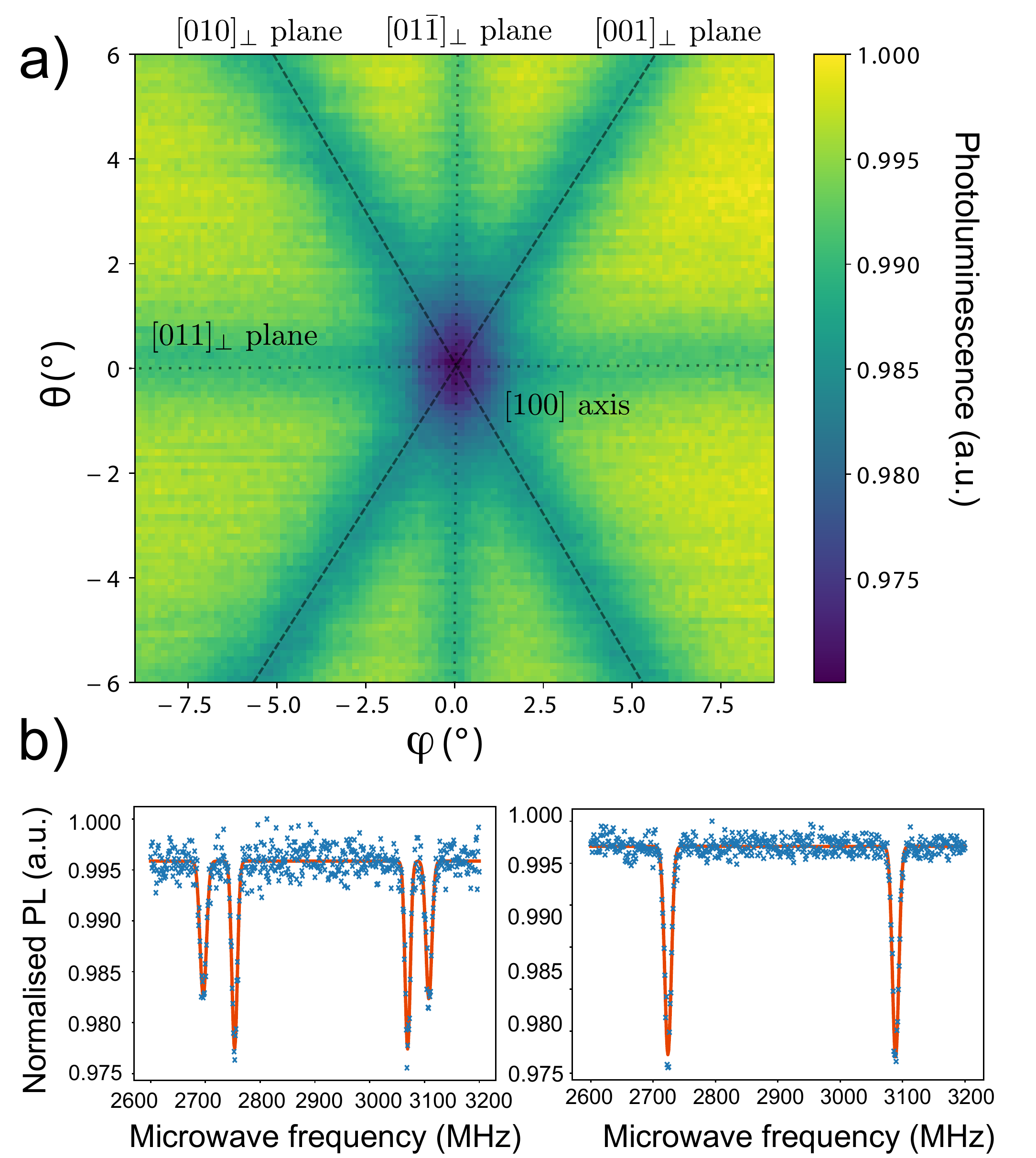}}
\caption{a) NV$^-$ photoluminescence as a function of the magnetic field angle around the [100] crystalline direction, at a fixed amplitude $|\vec{B}|=115$G. The planes orthogonal to the [010], [001], [011], and [01$\bar 1$] directions are indicated by dashed lines. b) Left: Optically-detected-magnetic-resonance (ODMR) spectrum taken at the angular position $(\phi, \theta)=(3^\circ,3^\circ)$. Right: ODMR spectrum taken at the exact center of the map, {\it i.e.} the [100] direction.}
\label{map}
\end{figure}

The negatively charged nitrogen-vacancy center has a zero-field splitting $D=(2\pi) 2.87$ GHz in the ground state, related to the dipole-dipole interaction between the spin of the two unpaired electrons. One of the most important properties of the spin of the NV$^-$ center is that it can be optically polarized and read-out and features long relaxation time ($\approx $ms) even at ambient conditions, which can be used to detect other species via CR and even to polarize them. 
Many paramagnetic defects in diamond also carry a spin 1.
The zero-field splittings are in fact a fingerprint of the defect. To measure it, one can tune the angle and magnitude of an external magnetic field angle to cause cross-relaxation with the NV centers. The obtained CR position can then be used to trace back the zero-field splitting of the defect.
Most experimental studies have been carried out using heavily doped crystals grown by the High-Pressure-High-Temperature (HPHT) process \citep{Epstein,Wang, armstrong_nvnv_2010, Hall, Wickenbrock, Alfasi}. 
Recent efforts in the doping processes have also made it possible to reach NV centers concentrations in the 3 to 5 ppm range in CVD grown samples \cite{Edmonds, TALLAIRE2020421, MINDARAVA2020182}, opening a path towards detecting other paramagnetic defects in CVD grown diamonds. 
The sample we use in this study was grown using CVD with the addition of 500ppm of N$_2$0 to the H2/CH4 (96/4) gas phase \cite{TALLAIRE2020421}.
Then high energy (10MeV) electron irradiation at a fluence of $2\times18$ cm$^{-2}$ and at a temperature of 900 degrees celsius was realised.
It was shown that in this sample the $T_2^*$ was not degraded after annealing, yet the NV density was large enough to enable concentration dependent longitudinal relaxation \cite{TALLAIRE2020421, jarmola_longitudinal_2015}.  
 
In this study, we use a homebuilt confocal microscope that comprises a 1mW green laser, an objective with a numerical aperture of 0.25 to focus the laser onto the sample as well as to collect the NV photoluminescence (PL). The PL was filtered from the green laser using a dichroic mirror and a notch filter. It was then coupled to a multimode fiber and detected by an avalanche photodiode. The magnetic field scans were realized using a C-shaped electromagnet driven by a current generator (HP 33120A).  We then monitor the NV PL synchronously with the changes in the magnetic field.
\begin{figure}[ht]
\centering
{\includegraphics[width=\linewidth]{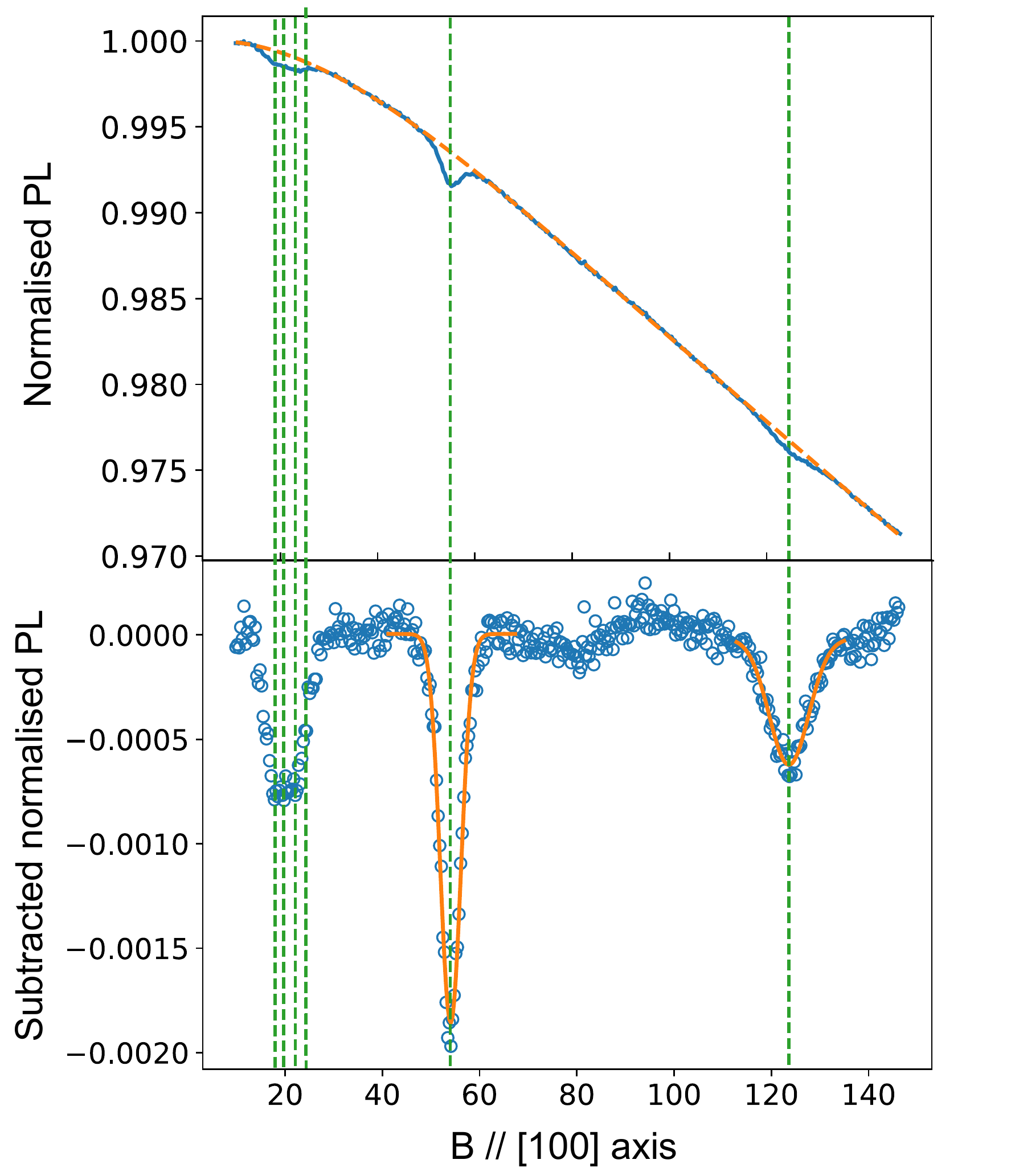}}
\caption{Optical detection of cross-relaxations. {Top}: NV$^-$ photoluminescence counts as a function of the magnetic field amplitude along the [100] crystalline direction (plain line) and fourth-order polynomial fit (dashed line). {Bottom} Curve obtained by subtracting the polynomial fit to the above signal (circles). The simulated cross-relaxations positions are shown by the green dashed, vertical lines. Gaussian fits to the second and third dips are shown by plain orange lines.}
\label{scan}
\end{figure}
Contrary to the more commonly employed $[ 111 ]$ direction, we scan the magnetic field along the diamond $[100]$ crystalline direction. 
Looking at the diamond structure (bottom right in Fig.  \ref{fig1})-a), it can be noticed that, in this direction, the four possible projections of the NV centers along the magnetic field are identical. This means that a B field scan along this direction cannot make the transition energies of the different classes of NV centers cross. Under arbitrary angles, this could otherwise result in several low field PL features \cite{van_oort_optically_1991, akhmedzhanov_magnetometry_2019} that could mask cross-relaxation from other species and also reduce the CR contrast (see section 4 of the supplementary materials for data showing scans along the [111] direction). 
The downside of this choice of the [100] direction is that at large amplitudes, the transverse component of the magnetic field depolarizes all classes of NVs, thus limiting the magnetic field range that can be employed and thus the range of defects that can be detected.

In order to identify the $[100]$ direction, we perform an angular scan of the magnetic field using a dual-axis goniometer (Thorlabs GNL20-Z8) that holds a permanent magnet at a fixed distance from the sample. 
Fig. \ref{map}-a) shows the NV photoluminescence as a function of magnetic field direction, referenced by azimuthal and polar angles $(\phi, \theta)$ with respect to the $[100]$ direction, using the above described CVD-grown sample. The PL is seen to drop for particular values of the magnet angular coordinates that correspond to specific crystalline axes.
Such a drop of the PL, also observed in
\cite{van_oort_optically_1991, van_oort_cross-relaxation_1989, jarmola_longitudinal_2015, akhmedzhanov_microwave-free_2017, akhmedzhanov_magnetometry_2019, holliday_optical_1989, choi_depolarization_2017}, correspond to situation where the transitions of NV centers become degenerate. 
The planes orthogonal to the [010], [001], [011], and [01$\bar 1$] directions are indicated by dashed lines and show the locus of the cross relaxation. These planes all cross on the [100] axis.
The origin of these sharp changes in the photoluminescence was attributed to cross relaxation between polarized NV centers and rapidly decaying NV centers, so called fluctuators \cite{choi_depolarization_2017}, the precise origin of the latter remains unknown.
The width of the CR features when the magnetic field crosses a plane perpendicularly was found to be compatible with the NV decoherence rate  ($\approx$ 6 MHz). The contrast is determined both by the fluctuating NV and polarized NV concentrations.

Such a map can in itself be useful for measuring magnetic fields without microwave \cite{akhmedzhanov_magnetometry_2019}. 
It also enables to identify the crystalline axes. 
To verify that the central line is the [100] axis, we realize microwave scans around the NV transitions. Away from the crystal planes, we consistently observe 8 ODMR lines coming from the four $| m_s=0\rangle$ to $| m_s=\pm1\rangle$ spin transitions of the \{111\}-oriented NV centers. On the planes orthogonal to the [010], [001], [011], and [01$\bar 1$] directions however, we expect degeneracies. 
Fig. \ref{map}-b) shows an ODMR spectrum taken at $(\phi, \theta)=(3^\circ,3^\circ)$. As expected, at this position, two pairs of NV classes cross.
Fig. \ref{map}-b) shows an ESR taken at the angle  $(\phi, \theta)=(0^\circ,0^\circ)$ showing only two features, as expected from a [100] axis. Using this goniometer, the magnetic field angle could be finely adjusted so that the four NV lines become fully degenerate along the $[100]$ direction with an error estimate of $\pm 0.5^\circ$.

Fig. \ref{scan} shows the PL as a function of magnetic field amplitude along the $[ 100 ]$ direction, in the 15~G to 145~G range. 
Calibration of the magnetic field amplitude was realized by applying microwave signals at varying frequencies in 2 MHz steps on several magnetic field scans. The microwave induced PL features in the scan are then used to relate the magnetic field to the applied voltage. 
Three features appear in the overall PL evolution in this spectrum : one feature at 20~G,  56~G and 122~G. 
For these data, the averaging was done for 24 hours, but the features appear already with a signal to noise ratio greater than 1 after 10 minutes.  
We also note an overall drop of the PL, as a result of state mixing in the optically excited state \cite{DOHERTY20131}. 
In order to let the three salient features detach better from the spectrum, we fitted a 4th-order polynomial to the PL evolution, without the spectral bumps, and subtract it from the data. The subtracted curve is displayed below.  The width and contrast of each peak can now be compared and their position can be determined with greater accuracy. 

\begin{figure}[htbp]
\centering
{\includegraphics[width=\linewidth]{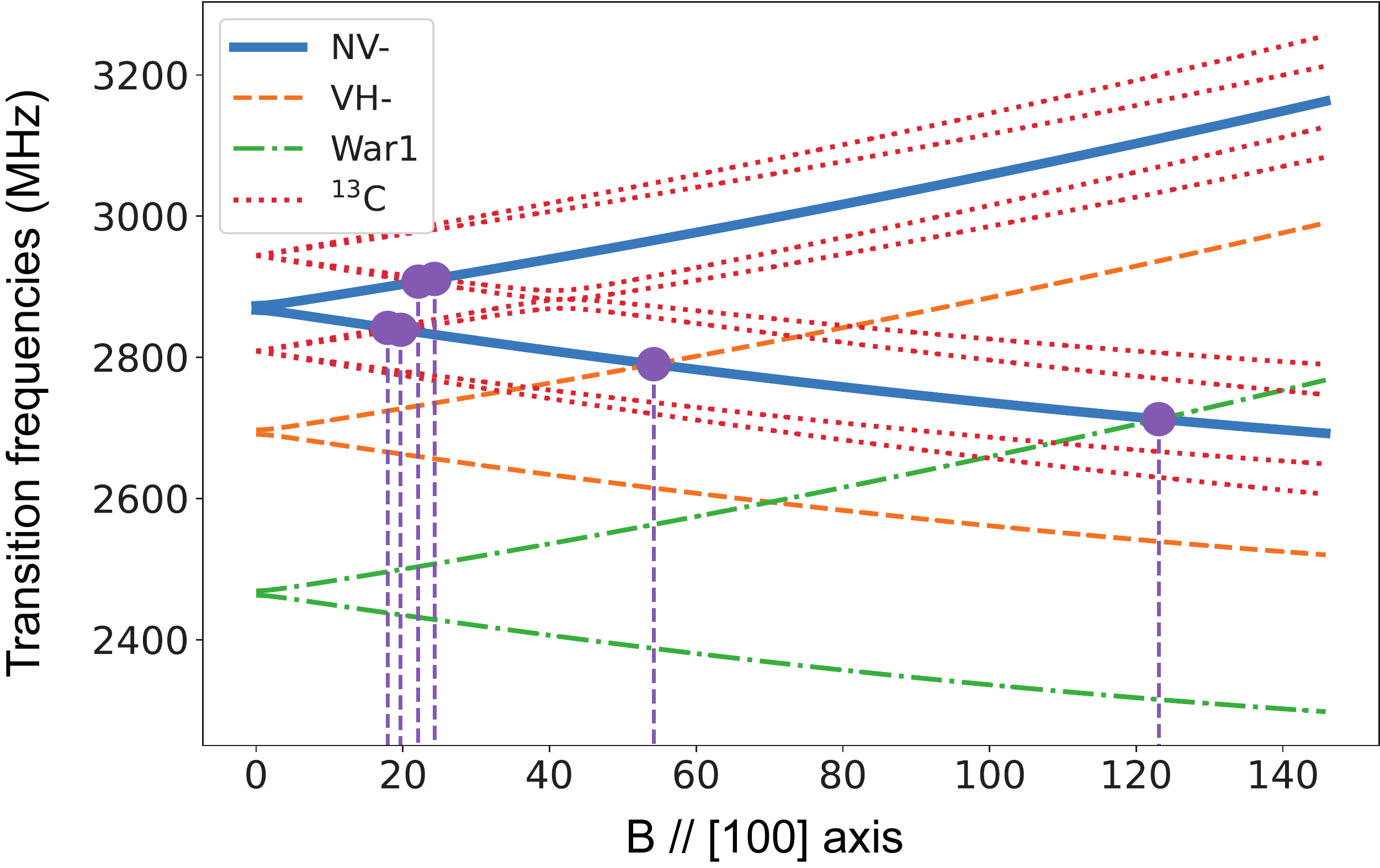}}
\caption{Simulated transition energies of the various considered spins as a function of a magnetic field aligned to the [100] axis. The NV centers electronic spin transitions are shown by the plain thick line, the VH$^-$ using dashed lines, the WAR1 using dash-dotted lines and $^{13}$C-NV pair using dotted lines. The amplitudes of the magnetic field where the energy level of the NV center crosses the one of another species are represented by vertical dashed lines.}
\label{energy-levels}
\end{figure}

In order to attribute the three features to their respective defects, we run a similar scan (Section 1 in the Supplementary Materials) on a type Ib electron irradiated HPHT diamond crystal with a NV concentration expected to be in the 5-20 ppm range, in the same $[100]$ direction. Only the first feature appeared. 
This observation guided us to search for CVD-related defects as good candidates for the last two features. Paramagnetic defects in CVD-grown diamonds have been extensively studied using EPR  \citep{Ashfold}. These studies demonstrate that hydrogen-related complex such as the NVH, VH, VH2 complexes can all be stable in diamond. Although their composition is not always known, several zero-field splittings $D$ can be found in the literature.
Given the small difference between the NV transitions and the observed features, we only concentrate on the reported defects with ZFS that are close to the NV$^-$.
The middle column of table \ref{table}, shows the zero-field splitting for two such spin-1 defects found in \citep{cruddace2007magnetic}, namely the negatively charged hydrogen-vacancy (VH$^-$) and WAR1 defects. The latter was analyzed in EPR but its exact structure is unknown \citep{cruddace2007magnetic}.
\begin{table}[htbp]
\centering
\caption{\bf Zero-field splitting parameter $D$ for the spin-1 species in our sample}
\begin{tabular}{ccc}
\hline
$D_z$ estimation (MHz) & Cruddace's work\citep{cruddace2007magnetic} & Our work \\
\hline
NV$^-$ & 2872(7) & * \\
VH$^-$ & 2706(30) & 2694(5)  \\
WAR1 & 2466(60) & 2470(10) \\
\hline
\end{tabular}
  \label{table}
\end{table}
Fig. \ref{energy-levels} shows the frequencies of the NV$^-$, the VH$^-$ and WAR1 spin transitions as a function of the magnetic field amplitude along the $[100]$ direction. The points were the NV levels cross the other defects can give rise to cross-relaxation.
Using this theoretical calculations, we find that the second peak in Fig. \ref{scan} coincides very well with a CR that would occur at the crossing between the $|m_s=0\rangle$ to $|m_s=-1\rangle$ NV transition and the $|m_s=0\rangle$ to $|m_s=+1\rangle$ VH$^-$ transition. 
A gaussian fit to the feature in Fig. \ref{scan} enables us to extrapolate a value for $D=2694(5)$MHz that coincides very well to that of the VH$^-$ within the error margins indicated in \citep{cruddace2007magnetic}.
The third peak at 122 G would correspond to a spin defect that has a $D=2470(10)$MHz that also matches well the one of the WAR1 defect. 
The values of $D$ for these two defects and the error bars from our measurements are now included in the table \ref{table}.
The error bars are estimated by taking into account the precision on the NV$^-$ ZFS, on the angle, on the magnetic field calibration and the precision of the fits. 
We obtain a factor of 6 improvement over the precision reported in \citep{cruddace2007magnetic}. We thus attribute the second and third feature to cross relaxation between the NV center and the VH$^-$ and WAR1 defect. Scans along the [111] direction shown in the Supplementary Material also corroborate this conclusion.

The remaining broader feature in the spectrum could be coming from dipolar coupling between the spin of NV centers and paramagnetic defects that are present also in HPHT-grown diamonds, such as the substitutional nitrogen, also called $P_1$ centers ($[P_1]\approx 5-20$ ppm in our sample) or the $^{13}$C atoms (natural isotopic abundance [$^{13}$C]= 1\%). 
Using the results from \cite{simanovskaia_sidebands_2013}, we first intended to correlate its spectral position to the transitions of dipolar-coupled NV and $P_1$ centers. 
 $P_1$ centers have a zero-field splitting of only 100 MHz. In order to cross one of the two NV spin transitions at a magnetic field of 20 G, they would then have to be coupled off-resonantly to the spin of nearby NV centers. The resulting pair could then be coupled resonantly to a nearby polarized NV.
The obtained eigenfrequencies of the dipolar coupled P$_1$-NV pair that we extracted were all inconsistent with the spectral positions of our observed peak within our spectral resolution error ($\approx$1G) (see Section 3 of the supplementary materials). This leads us to consider instead the nuclear spin of the $^{13}$C atoms as the most likely candidate.

The nuclear spin of the $^{13}$C atoms do not have a ZFS and its gyromagnetic factor is four orders of magnitude lower than that of the electron. 
Here again, in order to give a CR contribution to the spectrum at modest magnetic fields, the nuclear spin of the $^{13}$C would first have to be strongly coupled to an NV center. The resulting pair could then be co-resonant with the spin transitions of a bare NV center, as depicted in Fig. 1. This situation can manifest itself when the nuclear spin of $^{13}$C atoms are a few shells distant from the nitrogen-vacancy centers. A strong hyperfine coupling rate of 130 MHz can for instance be reached with a $^{13}$C that is only one shell away from the NV center. 
The modeling of this interaction is presented in the section 2 of the supplementary materials. The dotted lines in Fig. \ref{energy-levels} show the four transition frequencies of the first shell $^{13}$C coupled to an NV center. The crossings between the $|m_s=0\rangle$ to $|m_s=\pm 1\rangle$ NV transition and the four transitions of this  $^{13}$C-NV pair occur at magnetic field values around 20~G. 
We added these four lines to the experimental spectrum shown in Fig. \ref{scan}. A good agreement is found between the magnetic field at which of these four transitions cross the NV and the broad 20~G feature, letting us conclude that it is the result of a CR between an unpolarized $^{13}$C-NV pair and a polarized NV center.

In conclusion, we demonstrated all-optical detection of paramagnetic species in a diamond grown by the Chemical-Vapor-Deposition method. 
Using magnetic field scans along the [100] direction, we identified 3-body interactions between NV centers and $^{13}$C-NV pairs, and cross-relaxations between the spin of NV centers and the VH$^-$ and the WAR1 defects, by coupling them to NV centers. Our results offer prospects for more detailed studies of CVD-grown processes as well as for realizing quantum networks with these newly coupled spins.

\section*{Acknowledgments}
We would like to thank Neil Manson and Carlos Meriles for fruitful discussions. 
The authors would also like to thank SIRTEQ for funding. GH also acknowledges funding by the French National Research Agency (ANR) through the T-ERC project QUOVADIS.

\vspace{0.2in}

\section*{Appendix}
\subsection{Photoluminescence as a function of magnetic field in a HPHT sample}

\begin{figure}[htbp]
\centering
{\includegraphics[width=\linewidth]{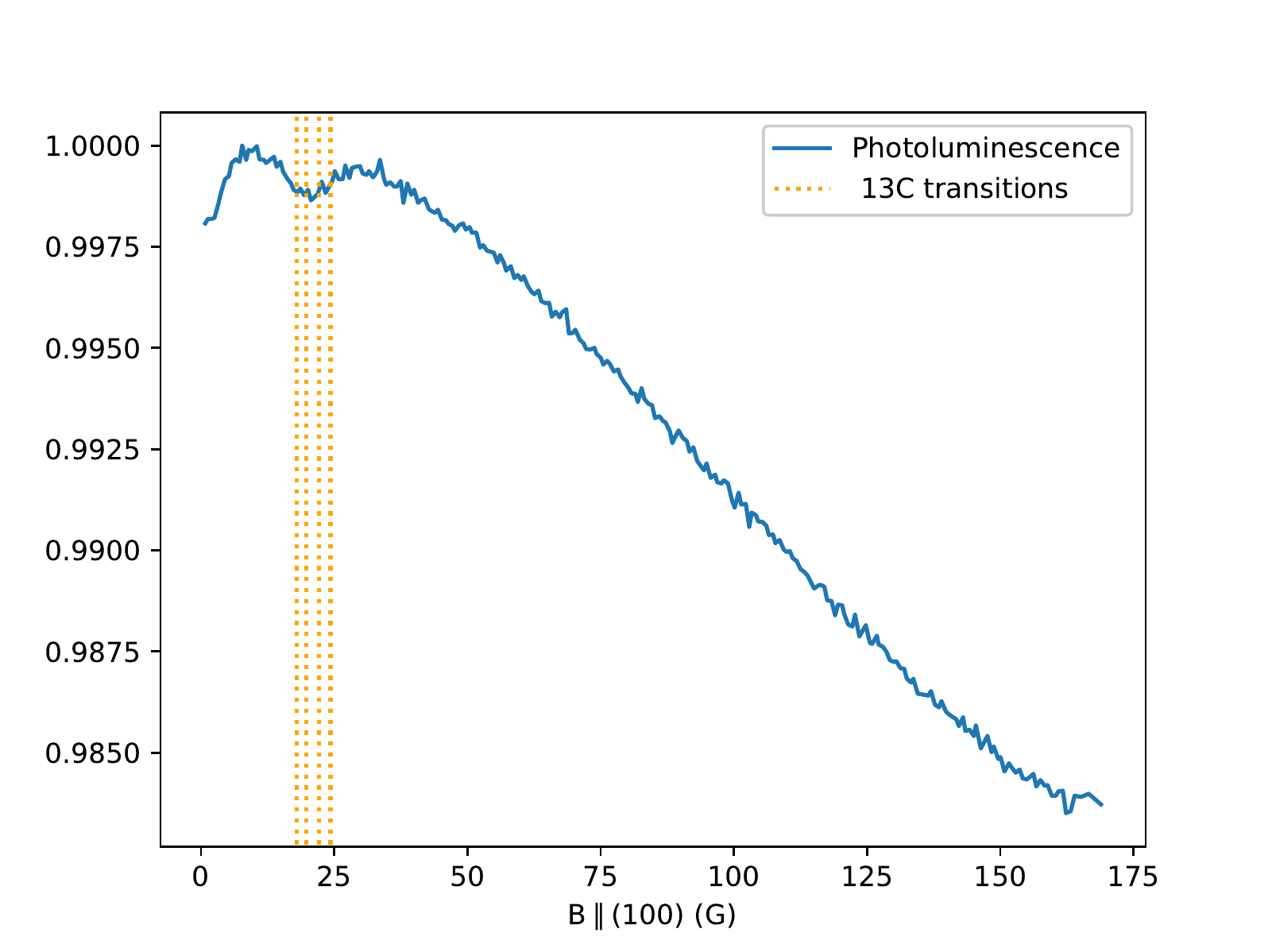}}
\caption{NV$^-$ photoluminescence rate while scanning the magnetic field along the [100] crystalline direction. Here a HPHT sample was employed.}
\label{fig1SI}
\end{figure}

The NV$^-$ photoluminescence as a function of the magnetic field amplitude that is presented in the Fig. 3 of the main text was realized on a CVD-grown sample. 
We also performed similar scans on several samples grown via the High-Pressure-High-Temperature (HPHT) method. 
Fig. \ref{fig1SI} shows a scan realized on HPHT samples that have been irradiated in order to increase the NV yield. 
The sample we use in Fig. \ref{fig1SI} is a Ib electron irradiated HPHT diamond crystal with a NV concentration expected to be in the 5-20 ppm range.
The PL feature that can be observed at 20G was also observed in the CVD-grown samples. We attributed it to the resonant dipolar interaction between a standard polarized NV center and one NV strongly coupled to the nuclear spin of a $^{13}$C atom sitting one shell away.
Notably, no other features could be detected at larger magnetic field values. We realized such scans on many different highly-doped HPHT samples and, systematically, no CR features are observed at 50 and 122 G. 

\subsection{NV-$^{13}$C Hamiltonian}

The full spin Hamiltonian of the NV-$^{13}$C complex can be written as follows : 
\begin{equation*}
\mathcal{H}=\mathcal{H}_{NV}+\mathcal{H}_{^{13}C}+\mathcal{H}_{HF},
\end{equation*}
where $\mathcal{H}_{NV}$ is the NV$^-$ spin Hamiltonian.
For one NV orientation, with quantization axis $z$, it reads 
\begin{equation}\hat{H}_{\rm NV}=\hbar D \hat{S}_z^2+ \hbar \gamma_e \bf B  \cdot \bf\hat S.
\end{equation}
$D$ is the zero-field-splitting and $\bf\hat S$ is the electronic spin operator.
$\mathcal{H}_{^{13}C}$ is the $^{13}$C nuclear spin Hamiltonian for a $1/2$ spin : $\mathcal{H}_{^{13}C}=\gamma_{n} B I_z$ where $\gamma_{n}=$10.7 MHz/T is the $^{13}$C gyromagnetic ratio, and $\mathcal{H}_{HF}$ is the hyper-fine interaction Hamiltonian : $\mathcal{H}_{HF}= \hat{\mathbf{S}}_{NV} \cdot \mathcal{A} \cdot \hat{\mathbf{I}}_C$.

In the case of a first shell $^{13}$C, the hyper-fine tensor $\mathcal{A}$ can be written as \citep{simanovskaia_sidebands_2013_SI} : $$ \mathcal{A} = \begin{pmatrix}
\mathcal{A}_{xx} & 0 & \mathcal{A}_{xz} \\ 0 & \mathcal{A}_{yy} & 0 \\ \mathcal{A}_{zx} & 0 & \mathcal{A}_{zz}
\end{pmatrix},$$
where $\mathcal{A}_{xx}=190.2(2)$ MHz, $\mathcal{A}_{yy}=120.3(2)$ MHz, $\mathcal{A}_{zz}=129.1(2)$ MHz, and  $\mathcal{A}_{xz}=\mathcal{A}_{zx}=-25.0(1)$ MHz. 

Diagonalizing the total Hamiltonian, we notice that the quantization axis of the nuclear spin (in the limit $\gamma_{n} B \ll A_{ij}$) is not the same in the manifold of the $\ket{m_s=0}$ state and in the $\ket{m_s=\pm 1}$ states \citep{alvarez_local_2015_SI}, meaning that the nuclear spin is not preserved by the electron spin flip, and gives rise to a splitting of the $\ket{m_s=0} \to \ket{m_s=\pm 1}$ transition in 4 distinct lines when the magnetic field is not aligned with the NV center \citep{jiang2018estimation_SI}.

\subsection{NV-P1 mutual spin flip transitions}

Transitions corresponding to a simultaneous flip of the NV$^-$ and P1 (substitutional nitrogen) spin states, mediated by the dipolar interaction between NV$^-$ and P1 electronic spins, are commonly observed on ODMR spectra \citep{simanovskaia_sidebands_2013_SI, kamp2018continuous_SI, alfasi2019detection_SI, lazda2020cross_SI}. Here, we investigate whether the cross-relaxation processes we observe could be partly due a three-body interaction between two NV centers and one P1 center.

In order for this process to be energy conserving, the P1 transition frequency has to match the energy difference between the $\ket{m_s=0} \to \ket{m_s=-1}$ and $\ket{m_s=0} \to \ket{m_s=+1}$ transitions, giving us the equation \begin{equation}
\label{eq_P1}
\nu^i_{P1}(B)=\nu^{0 \to +1}_{NV}(B)-\nu^{0 \to +1}_{NV}(B),
\end{equation}
where $\nu^i_{P1}$ is a transition between any of the P1 spin Hamiltonian eigenstates. Note that since we are scanning the magnetic field in the crystalline [100] direction, we do not have to take into account the four classes of NV centers and P1.

The P1 spin states are defined by its electron 1/2-spin and its nuclear 1-spin, giving a manifold of 6 spin states. When considering any possible transitions between the 6 eigenstates of the P1 Hamiltonian, we have therefore a total of 15 possible transitions, including the nuclear-like transitions which have been observed through mutual spin flip with NV centers \citep{alfasi2019detection_SI}. 

\begin{figure}
\includegraphics[width=\linewidth]{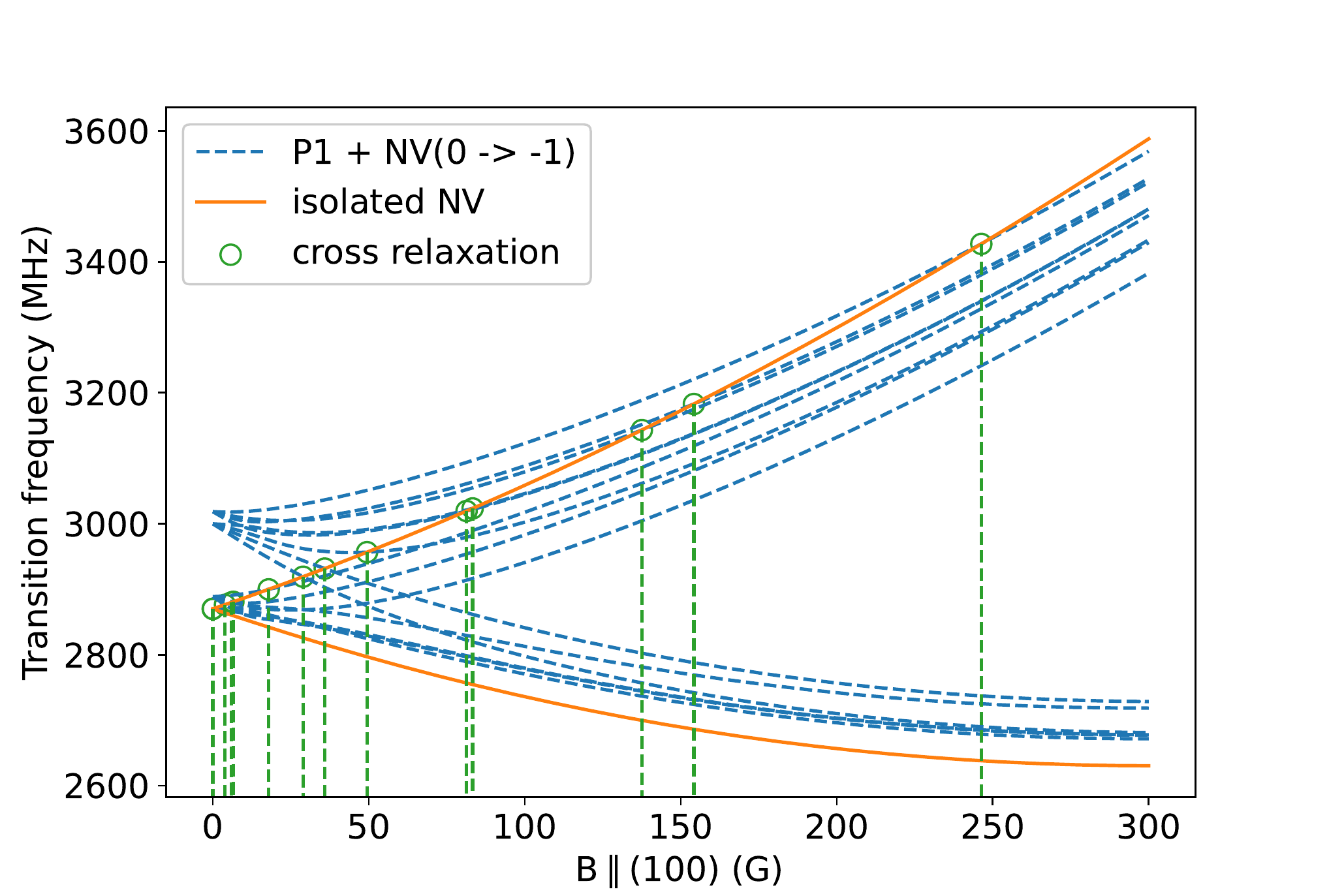}
\caption{Cross-relaxation condition for mutual flip between NV and P1 centers. The orange plain lines correspond to the NV center $\ket{m_s=0} \to \ket{m_s=-1}$ and $\ket{m_s=0} \to \ket{m_s=+1}$ transition frequencies as function of a magnetic field along the [100] crystalline axis, the blue dashed lines correspond to the frequencies of the 15 theoretical transitions of the P1 centers, added to the frequency of the NV $\ket{m_s=0} \to \ket{m_s=-1}$ transition. The green circles correspond to the particular magnetic fields where Eq. \ref{eq_P1} is verified, meaning where the energy of the P1 transition matches the energy difference between the two NV transitions.}\label{fig_P1}
\end{figure}

Using the numerical values of the P1 Hamiltonian from \citep{lazda2020cross_SI}, we solve Eq. \ref{eq_P1} for all possible P1 transitions (see fig. \ref{fig_P1}) therefore giving us all the possible magnetic field amplitudes in the [100] direction where we could observe NV-P1 mutual spin flip transitions. The predicted magnetic field values are 0, 3.89, 5.96, 6.58, 17.90, 28.93, 35.87, 49.44, 81.33, 83.28, 137.52, 154.20 and 246.34 G. None of these values correspond to any feature in our scans. We then concluded that the features we observed were not due to NV-P1 mutual flips.

\bibliographystyle{plain}

\subsection{Magnetic field scan along the [111] direction}
\begin{figure}
\includegraphics[width=\linewidth]{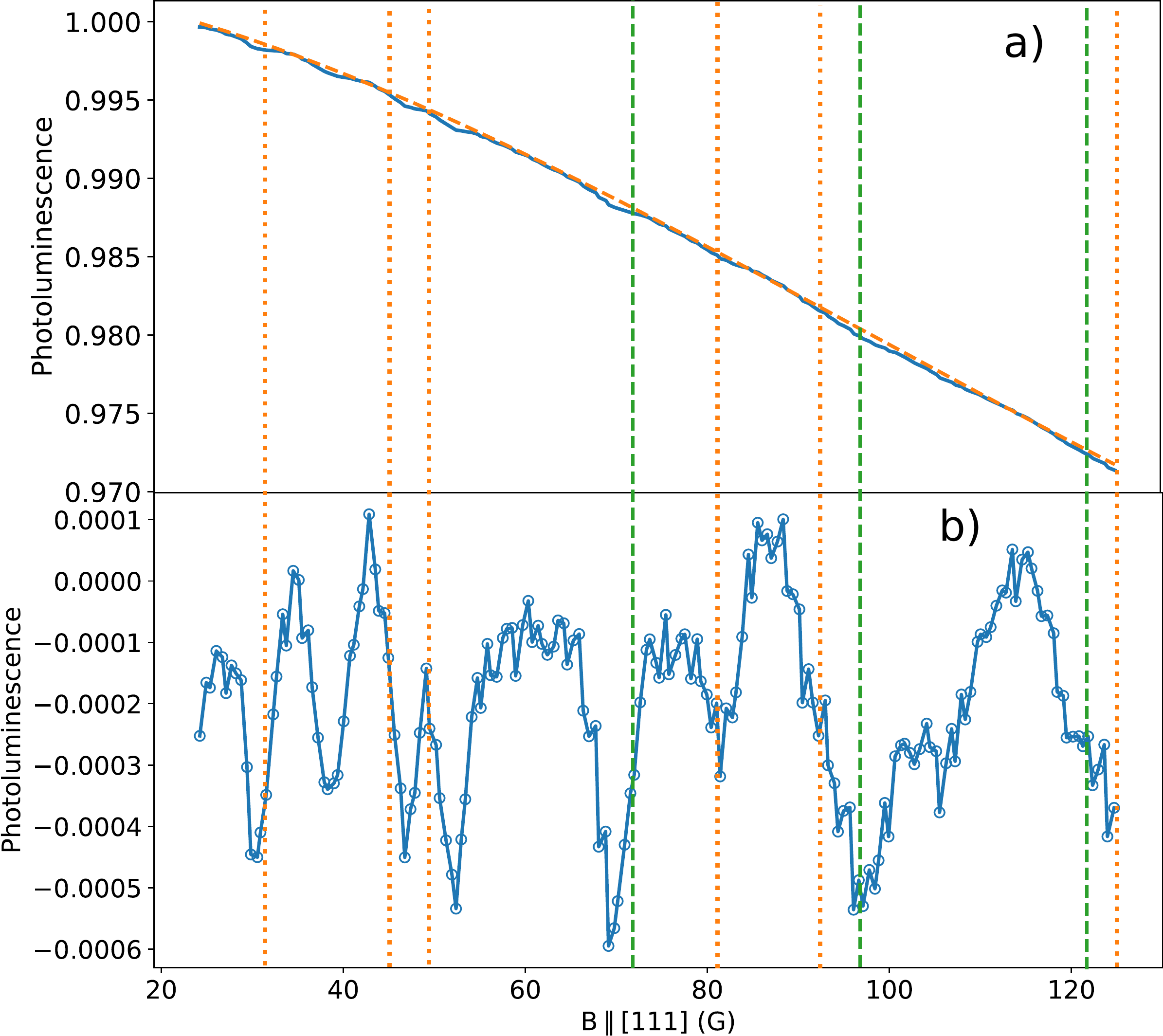}
\caption{\textbf{a)} Plain blue line : Photoluminescence as a function of the magnetic field amplitude in the [111] crystalline direction. Dashed orange line : 4th order polynomial fit of the decreasing envelope. \textbf{b)} Subtraction of the previous signal by the envelope fit. The vertical dotted orange lines correspond to the expected cross-relaxations with VH$^-$ and the vertical dashed green lines to the expected cross-relaxations with WAR1.}\label{scan_PL}
\end{figure}

\begin{figure}
\includegraphics[width=\linewidth]{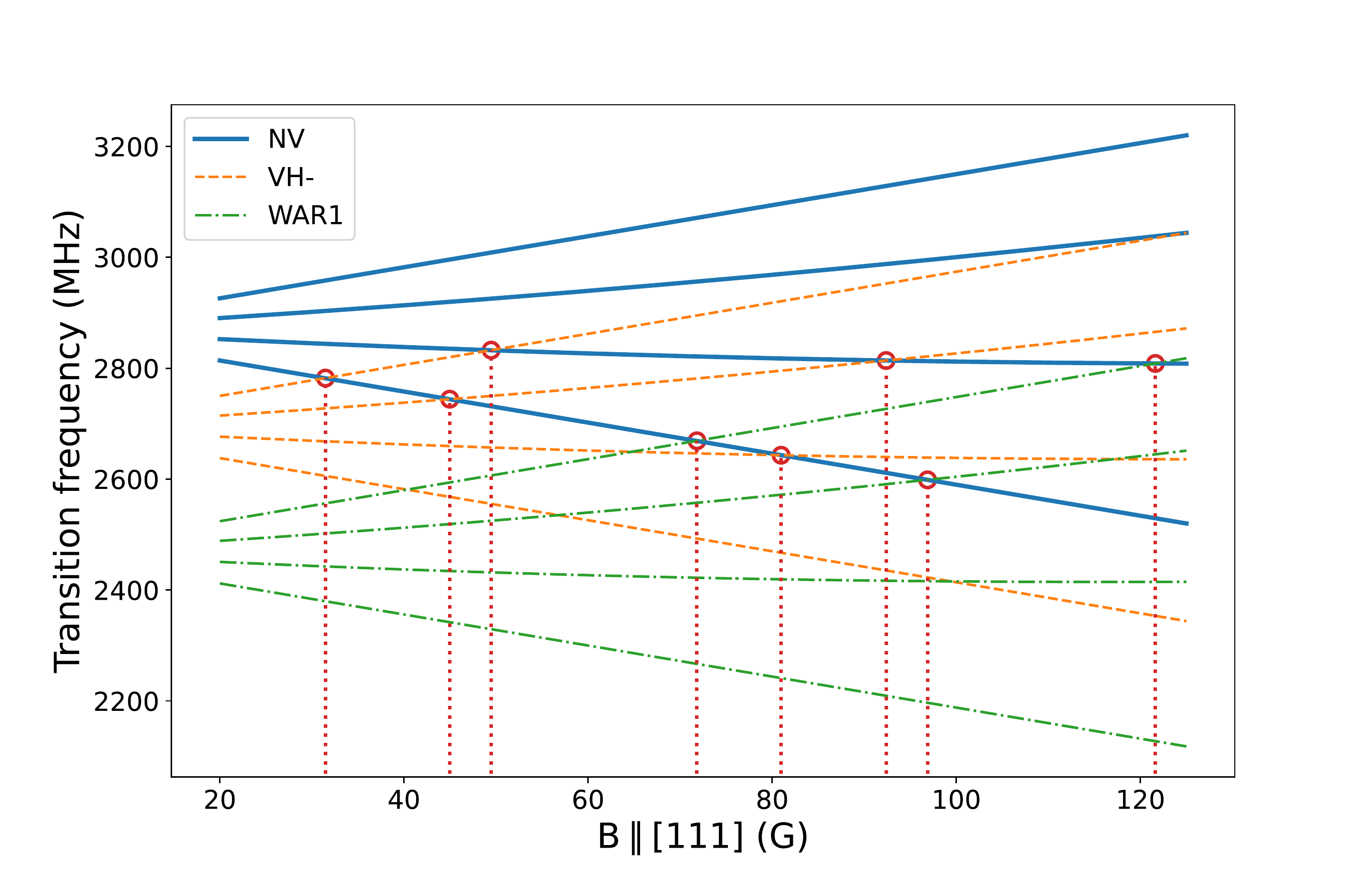}
\caption{Predicted transition frequencies of NV$^-$ centers (plain lines), VH$^-$ (dashed lines) and WAR1 (dash-dotted lines) when the magnetic field is scanned in the [111] direction. The magnetic field values for which  a transition of the NV center crosses one of VH$^-$ or WAR1 are also shown in Fig \ref{scan_PL}.}\label{transis_VHWAR}
\end{figure}

\begin{figure}

\includegraphics[width=\linewidth]{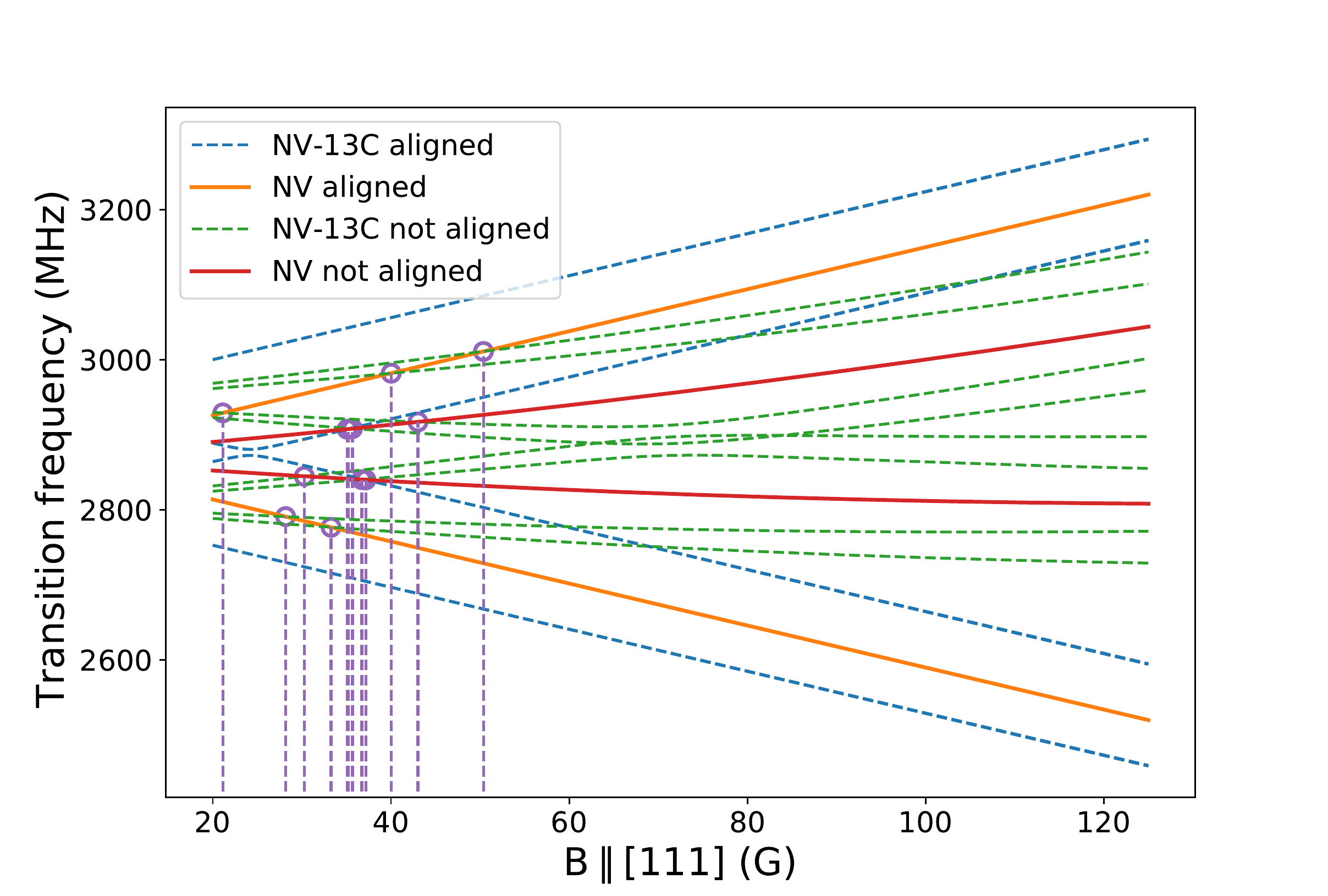}
\caption{Predicted transitions frequencies for NV centers (plain lines) and NV-$^{13}$C complex when the magnetic field is scanned in the [111] direction. Vertical lines represent the various degeneracy conditions.}\label{transis_13C}
\end{figure}

We have performed similar magnetic field scan as presented in Fig.3 of the main text, but scanning the magnetic field in the crystalline [111] direction instead of the [100], meaning that this time one class of NV centers is aligned with the magnetic field while the three others are misaligned and degenerate. The results in term of cross-relaxations generally means 4 times as many degeneracy conditions as the [100] case, since the two spins involved in the cross-relaxation process now have two possible orientations, and therefore two separate transitions.

It is then significantly harder to find cross-relaxations signatures, since they are both more numerous and with a smaller contrast than in the [100] case. The experimental results are shown in Fig. \ref{scan_PL}. Similarly to Fig. 3 of the main text, we have subtracted the envelope associated with the transverse field in order to better see the fine features of the scan. The predicted cross-relaxation conditions between NV centers and VH$^-$ or WAR1 are presented in fig. \ref{transis_VHWAR} and shown in Fig. \ref{scan_PL}. While the predicted values do not exactly match the experimental results, it seems like every predicted transition falls in the vicinity of an experimental line. Importantly, it should be noted that, unlike Fig.3 of the main text, the magnetic field was not calibrated with the lengthy NV ODMR at every field value in this case. Instead we have simply applied a homotethy on the current intensity in the electromagnet to match the field values at the maxima of the scan. This means that the magnetic field of the points of low field value in particular are more susceptible to the non-linearities due to the electromagnet hysteresis cycle.

Finally we have plotted in Fig.\ref{transis_13C} the expected cross-relaxations between NV and NV-$^{13}$C complex. It seems that these transitions are too diluted to appear in our experimental scan. The peak at $\sim$37 G in fig.\ref{scan_PL} could come from the resonance between NV and NV-$^{13}$C since they seem particularly dense at this particular field values.
It is indeed  the only peak that is not matching any of the NV-VH$^-$ or NV-WAR1 resonance.

\end{document}